# Tapered optical fibers as tools for probing magneto-optical trap characteristics

Michael J. Morrissey,[1,2] Kieran Deasy,[1,2] Yuqiang Wu (邬宇强),[2,3] Shrabana Chakrabarti,[2,a] and Síle Nic Chormaic[2,3]

[1] Department of Applied Physics and Instrumentation, Cork Institute of Technology, Cork, Ireland

[2] Photonics Centre, Tyndall National Institute, University College Cork, Prospect Row, Cork, Ireland

[3] Physics Department, University College Cork, Cork, Ireland

We present a novel technique for measuring the characteristics of a magneto-optical trap for cold atoms by monitoring the spontaneous emission from trapped atoms coupled into the guided mode of a tapered optical nanofiber. We show that the nanofiber is highly sensitive to very small numbers of atoms close to its surface.  The size and shape of the MOT, determined by translating the cold atom cloud across the tapered fiber, is in excellent agreement with measurements obtained using the conventional method of fluorescence imaging using a CCD camera. The coupling of atomic fluorescence into the tapered fiber also allows us to monitor the loading and lifetime of the trap. The results are compared to those achieved by focusing the MOT fluorescence onto a photodiode and it was seen that the tapered fiber gives slightly longer loading and lifetime measurements due to the sensitivity of the fiber, even when very few atoms are present.

---

[a] Current Address: Institut für Angewandte Physik, Technische Universität Darmstadt, D-64289 Darmstadt, Germany.

## I. Introduction

The fabrication of tapered optical fibers[1-4] has achieved renewed interest in recent years due to the wealth of applications for which they can be used, including super-continuum generators,[5] optical filters,[6] coupling devices into optical resonators,[7-10] and detection systems for gas[11] and surface adsorption.[12] Tapered fibers are fabricated by heating and pulling single mode fiber such that the fiber waist diameter is reduced to a size smaller than the propagating radiation wavelength, $\lambda$. This results in an increase in the transversal dimension of the fundamental mode, primarily in the evanescent field, and the dimension reductions can be achieved with minimal loss.[3] More recently, sub-micron tapered optical fibers, i.e. tapered nanofibers (TNF), are finding applications in laser cooling and trapping experiments,[13,14] with the goal of using the evanescent field of the TNF to trap and manipulate cold atomic samples outside the fiber,[15,16] as a possible alternative to current-carrying wire atom guides.[17-19]

Sagué et al.[20] have proposed an alternative trapping scheme using TNFs, whereby two-mode interference of blue detuned light is used to generate an array of microtraps close to the fiber surface. The advantage of such a trapping scheme lies in the fact that the TNF can be used to confine, manipulate, and probe cold atoms in a controlled manner. Such techniques could prove promising for positioning single particles (e.g. atoms, molecules, ions or biological samples) with high precision and good efficiency.

In order to fully comprehend the processes involved when a cold atom is close to the surface of a TNF it is imperative that a thorough understanding of the spontaneous emission rate of atoms located near the fiber surface, and the subsequent coupling into the guided modes of the fiber, be acquired.[21] Experimental[13] and theoretical[22,23] studies have shown that the spectral line shape is strongly influenced by the presence of the fiber due to the van der Waals and Casimir-Polder interactions that cause a shift to the atomic transition frequency. In the work presented here we utilize the coupling of the spontaneous emission from laser-cooled rubidium atoms to the guided modes of a tapered nanofiber to experimentally demonstrate the potential TNFs offer as highly efficient tools for probing MOT characteristics such as lifetime, loading, and cloud profile even when the number of trapped atoms is very small.

## II. Experimental Setup

### A. Tapered optical fibers

There are many techniques used to produce TNFs.[1-4] In our work, the TNFs are fabricated in a clean environment using a heat-and-pull technique. An oxy-butane flame is used to heat a stripped area of 780 nm single mode optical fiber, and two translation stages are used to simultaneously pull the fiber in opposite directions during the heating process. This has the effect of reducing the fiber diameter as a function of pull length and hot-zone size according to the adiabatic condition.[24] Probe light is passed through the fiber during the heat-and -pull process, allowing us to monitor the transmission through the fiber at all times. The TNFs may be imaged using a scanning electron microscope and, typically, they are smooth, have no evidence of surface defects, and a sub-micron waist.

We routinely fabricate TNFs with a diameter as small as 400 nm and a transmission greater than 90%. For the current experiments, we use a TNF with a diameter of 600 nm and a transmission of 85%, prior to installation in the vacuum chamber. For the purpose of redundancy, two similar TNFs are mounted vertically in the vacuum chamber and fed in and out of the chamber using a Teflon® fiber feedthrough system.[25] Fig. 1 is a schematic of the mounting system within the UHV chamber and illustrates the principle behind the measurements. During the preparation and installation of the fiber, the transmission decreased to 66%. This reduction in transmission may be due to the UV glue binding the fiber to the mount or dust particles falling on the fiber during the mounting process.

## B. MOT Setup

A standard MOT design[26-29] is used to produce a cloud of $^{85}$Rb atoms in the vicinity of the TNF waist. The cooling laser is locked 12 MHz red detuned from the $5S_{1/2}$, F = 3 → $5S_{3/2}$, F' = 4 closed cycle transition using Doppler-free saturated absorption spectroscopy. A repumper laser is tuned to the $5S_{1/2}$, F = 2 → $5S_{3/2}$, F' = 3 transition. Each cooling beam has an intensity of 2.4 mW/cm$^2$ and a diameter of 16 mm. The center point of a quadrupole magnetic field, created by a pair of anti-Helmholtz coils, is overlapped with the intersection point of the laser beams. The magnetic field gradient at the center of the trap is approximately 10 G/cm, with a coil current of 4 A. The background Rb vapor is controlled by resistively heating a Rb dispenser. The MOT is viewed orthogonally using CCD cameras and, with the aid of pairs of smaller coils placed around the chamber, the MOT is spatially overlapped with the waist region of the TNF. Once overlapped, the MOT has a cigar shape with a 1/*e* vertical length of 2 mm and a horizontal length of 1.3 mm, with an average MOT density of $4 \times 10^6$ atoms/mm$^3$.

## C. Fluorescence detection

Atoms trapped close to the surface of the TNF absorb light from the cooling beams and, subsequently, reemit light that couples into the guided modes of the fiber. A single photon counting module (SPCM, Perkin Elmer, SPC-AQRH-14-FC) is attached to one end of the fiber. As each reemitted photon is detected by the SPCM a TTL pulse is outputted to a counter (Hamamatsu Counting Unit, C8855) and the data is transferred to a PC, where it is displayed for a user-defined gating time.

## III. RESULTS AND DISCUSSION

## A. Fluorescence coupling efficiency

Before we could proceed to characterize the cloud of cold atoms using the TNF, it was essential that we determine whether fluorescence from trapped atoms was efficiently coupled into the tapered fiber. The demonstration was performed in three stages, while monitoring the number of photon counts on the SPCM. First off, the repumping laser was switched on, followed by the cooling laser and, finally, the magnetic field was switched on. This enabled us to distinguish between background noise from the repumping and cooling lasers and fluorescence emitted by the trapped atoms. Fig. 2 shows the number of photon counts per second detected for each of the three stages and, also, for the reverse sequence where field coils, cooling laser and repumping laser were switched off. The initial count rate of $1.5 \times 10^5$ $s^{-1}$, when all lasers and the magnetic field are off, is due to the dark count of the detector, and, to a greater extent, the various instruments and light sources in the laboratory. An increase of $2 \times 10^4$ $s^{-1}$ is observed when the repumping laser beam is turned on, followed by an increase of $4\times10^4$ s$^{-1}$ when the cooling laser beams are switched on. Once the magnetic field is switched on the MOT loads and a dramatic increase of $4\times10^5$ s$^{-1}$ is observed from the trapped atom fluorescence.

The count rate detected by the SPCM, $\eta_p$, can be determined using the following formula[13]

$$\eta_p = N_{eff}\eta_f\gamma_{sc}\eta_D T, \qquad (1)$$

where $N_{eff}$ is the effective number of atoms, $\eta_f$ is the coupling efficiency of spontaneous emission into the guided modes of the fiber, $\eta_D$ is the quantum efficiency of the detector, and $T$ is the transmission through the TNF from its waist region to one end. The atomic scattering rate, $\gamma_{sc}$, can be written as a function of laser intensity and detuning of the cooling laser from the cooling transition. For our experimental conditions, $\gamma_{sc} = 6.5\times10^5$ s$^{-1}$. To determine $N_{eff}$, we assume that any atoms beyond a distance of 300 nm from the surface of the fiber have very little effect on the spontaneous emission detected by the photon counter. By overlapping this hollow cylinder shaped observation area with the average density of the MOT, the effective number of atoms contributing to the spontaneous emission detected by the SPCM was calculated to be six per second. This leads to a resultant count rate of $3.74\times10^5$ s$^{-1}$, which is in very good agreement with the results shown in Fig. 2.

The MOT profile can be determined by moving the cloud of cold atoms across the waist of the tapered fiber and measuring the photon count rate as a function of fiber position. Due to laser beam alignment and anti-Helmholtz coil arrangement, the MOT is initially formed several mm away from the TNF. A single magnetic coil is used to translate the MOT across the TNF in the horizontal direction. A CCD camera combined with a telescope system was used to record the movement of the MOT. The imaging system is calibrated in terms of distance and permits us to convert time into the distance moved by the atomic cloud. When the edge of the cloud begins to overlap the TNF, fluorescence emitted by cold atoms close to the fiber surface couple into the fiber. As the center of the MOT approaches the TNF waist the fluorescence count rate increases with the increase in atomic density. The opposite effect is observed as the center of the atom cloud passes the TNF waist. Recording the photon count rate at each position of the atom cloud enables us to determine the cloud shape and atom density. Fig. 3 is a plot of photon count rate as the cloud is moved across the waist region of the fiber. For the sake of comparison, we have also determined the cloud profile using standard techniques, whereby the distribution of atoms in the MOT was estimated by extracting a frame from a video which records the fluorescence from the cloud. The relative intensity of the pixels can be determined from the extracted frame and a cross-section, taken perpendicular to the TNF, gives the spatial distribution of the atomic cloud along the axis of motion. These results are also shown in Fig. 3 and the signal amplitude is normalized to the maximum count rate of the SPCM. From a Gaussian fit the $1/e$ diameter of the atom cloud was found to be 1.31 mm for the SPCM plus TNF method and 1.27 mm for the image cross-section method, showing good agreement between both techniques.

## B. MOT loading

The loading of the MOT as a function of time was next determined and a comparison made between the results using the SPCM plus TNF system versus the standard photodiode imaging scheme. The MOT is loaded from background Rb vapor in the UHV chamber, produced by resistively heating a Rb dispenser. The number of atoms in a MOT is proportional to the intensity of fluorescence emitted from the trapped atoms. The first technique we used to measure the number of atoms in the MOT as a function of time involved focusing the emitted fluorescence from the entire MOT onto a photodiode which measures the total optical power. The resultant photocurrent was converted into a voltage signal using a load resistance and sent to a PC. By monitoring this voltage, a real-time loading evolution of the MOT was inferred. The second technique we used relied on the fact that the loading of atoms in the center of the MOT, around the tapered fiber waist, is a function of the loading of the entire MOT. Therefore, by monitoring the fluorescence from the center of the MOT coupled into the TNF, the loading rate of the entire atomic cloud can be deduced by determining the count rate on the SPCM.

For each of these two techniques, the Rb dispenser was switched on and allowed to stabilize for 15 minutes prior to data recording, and the MOT was loaded while ensuring it overlapped the TNF waist at all times. The resultant loading curves are shown in Fig. 4. For the measurements taken using the photodiode, background fluorescence had to be compensated for, whereas the TNF has the advantage that it is insensitive to the background signal since only fluorescence from atoms very close to the fiber can couple into it. As can be seen from Fig. 4, the results obtained from the TNF are in reasonably good agreement with the results achieved using the photodiode imaging technique, indicating that the TNF is indeed a good tool for measuring such MOT parameters. The $1/e$ loading time was determined to be 0.43 s from the photodiode data and 0.51 s from the SPCM data. Note that we consistently obtained longer loading times using the tapered optical fibre than when using fluorescence imaging. This discrepancy is attributed to the MOT behavior switching from an initial temperature limited regime for few atoms, to a constant density regime once the number of trapped atoms is greater than $\sim 5 \times 10^4$. However, more detailed study on this behavior is necessary in order to fully determine the cross-over between both regimes.

## C. MOT decay

To measure the MOT lifetime the Rb dispenser was switched off, leading to a decrease in the atom capture rate into the trap. As a consequence, the MOT population decays due to the loss rate from the trap exceeding the capture rate, and the trap lifetime can be determined by monitoring this decay. The same two techniques as used for the trap loading characteristics were used to determine the trap lifetime: (i) the fluorescence was focused onto a photodiode and (ii) the fluorescence from atoms trapped close to the fiber waist and coupled into the TNF was measured using the SPCM attached to one end of the fiber.

Fig. 5 shows the lifetime evolution of the trap population once the Rb dispensers are switched off. It can be seen that the SPCM plus TNF technique yields a longer 1/*e* lifetime measurement of 13 s as compared to 9.4 s obtained using the photodiode imaging system. This effect can be explained if one assumes that the center of the MOT does not decay as fast as the outer regions of the MOT since the hotter atoms escape from the MOT quicker than colder atoms at the center region. Also, when the number of atoms remaining trapped in the MOT is relatively small the photodiode is less sensitive to such low levels of fluorescence, whereas the TNF remains highly sensitive. This shows a clear advantage of using the tapered nanofiber in a very low density MOT or for detecting very few atoms compared to standard photodiode imaging systems.

Once more, the discrepancy observed between both techniques could be due to the fact that the TNF is sensitive to the local atomic density at the centre of the atom cloud through which it passes, whereas the photodiode is sensitive to overall density changes within the MOT. In the constant-density regime, the atom cloud may be characterized by an elongated MOT symmetry with a flat-topped profile, the atom density being distributed uniformly along the elongated axis (i.e. along the length of the TNF).[30] In this regime a repulsive force exists between atoms due to multiple photon scattering, resulting in an additional heating effect of the cloud. Any further loading of atoms into the cloud causes the cloud diameter to increase, thereby maintaining a constant density. Conversely, as the MOT begins to decays, the hotter atoms at the outer edges of the cloud escape first, reducing the diameter of the MOT and, again, maintaining a constant density profile in the center of the cloud. As a result, the density at the center of the cloud appears to decay slower than that at the outer edges. The decay rate detected by the SPCM, therefore, appears slower than that of the photodiode.

Once the number of trapped atoms decreases below $\sim 5\times 10^4$, the cloud enters the temperature limited regime and the density profile follows that of a Gaussian distribution. In this regime the atoms have a temperature equivalent to those in a molasses setup with the same beams parameters.[31,32] As more and more atoms escape from the MOT the diameter of the cloud remains constant[30,33] and, thus, the density decreases linearly with the number of trapped atoms.[34] In this case we would expect that the photodiode and the SPCM would measure approximately the same decay rates and further studies are required in order to quantify this behavior.

## IV. CONCLUSIONS

In conclusion, we have experimentally demonstrated that a tapered optical nanofiber is a very efficient tool for monitoring and probing fluorescence emitted by atoms trapped in a magneto-optical trap and located in the vicinity of the fiber waist. Through these measurements we have shown that the optical nanofiber can be used to determine MOT characteristics such as the cloud profile, loading time and lifetime. A very good signal to noise ratio was observed since the fluorescence is preferentially coupled into the fiber guided mode. Although it appears that the signal-to-noise ratios of both detection systems are of the same magnitude (c.f. Figs. 4 and 5), these variations are, in fact, due to intensity fluctuation within the MOT and, therefore, appear on both detectors with the same relative amplitudes. In practice, the SPCM has improved signal-to-noise over the photodiode.

When the results using the TNF are compared to the measurements taken using standard photodiode imaging techniques, both sets of results were largely found to be in good agreement. While the lifetime determined using the optical nanofiber was longer than that using the photodiode imaging, and can be explained by the MOT switching from the temperature limited to the constant density regime, the discrepancies in the loading times, though smaller, cannot be fully explained using this argument. To study this effect in more detail further work will concentrate on increasing the MOT loading time by reducing the background vapor pressure and reconfiguring the MOT so that the minor axis of the cloud lies along the axis of the optical fiber. This should enable us to systematically study the difference between loading times for the two methods and its dependency on the cloud profile. Insightful investigations into density distributions in magneto-optical traps as a function of the number of atoms trapped should be possible with a high degree of precision. A further goal of this work will be to study the cloud profile for different MOT beam detunings and intensities in order to establish the cross-over region between the Gaussian and the flat-topped profile as discussed by, amongst others, Arnold and Manson.[35]

Our results indicate that the TNF may be a preferential technique for determining MOT characteristics when dealing with very small numbers of atoms. These results show that not only do TNFs have the possibility of trapping and manipulating cold atoms but also demonstrate that nanofibers can act as a probing tool for various parameters associated with clouds of cold atoms. This technique should also facilitate studies on the influence of van der Waals and Casimir-Polder surface interactions on the line shape of the fluorescence emitted from trapped atoms and on using the optical nanofiber as a manipulation tool for cold and ultracold atoms. For example, a tapered optical fiber could be used to transfer atoms into a series of Fresnel-based microtraps.[36] These techniques could also be extended to probing and manipulating other systems, including molecules and ions.

## V. ACKNOWLEDGEMENTS

This work is supported by Science Foundation Ireland under Grant Nos. 02/IN1/128 and 07/RFP/PHYF518. KD and YW acknowledge support from IRCSET under the Embark Initiative.

# FIGURE CAPTIONS

FIG. 1. (Color online) Schematic of the mount for the TNF in the UHV chamber. The expanded inset represents the trapped atoms around the waist of the nanofiber that spontaneously emit radiation that is coupled into a guided mode of the fiber.

FIG. 2. (Color online) Measurement of photon count rate coupled into the TNF. The large increase in the count rate when the magnetic field is switched on is due to atoms trapped near the fiber fluorescing and the radiation being coupled into the fiber guided modes.

FIG. 3. (Color online) Estimation of atom cloud profile using the SPCM plus TNF system (red curve) and an image cross-section (blue curve).

FIG. 4. (Color online) Loading of the MOT as observed by the photodiode (blue curve) and the SPCM plus TNF system (red curve).

FIG. 5. (Color online) Lifetime of the MOT as observed by the photodiode (blue curve) and the SPCM plus TNF system (red curve).

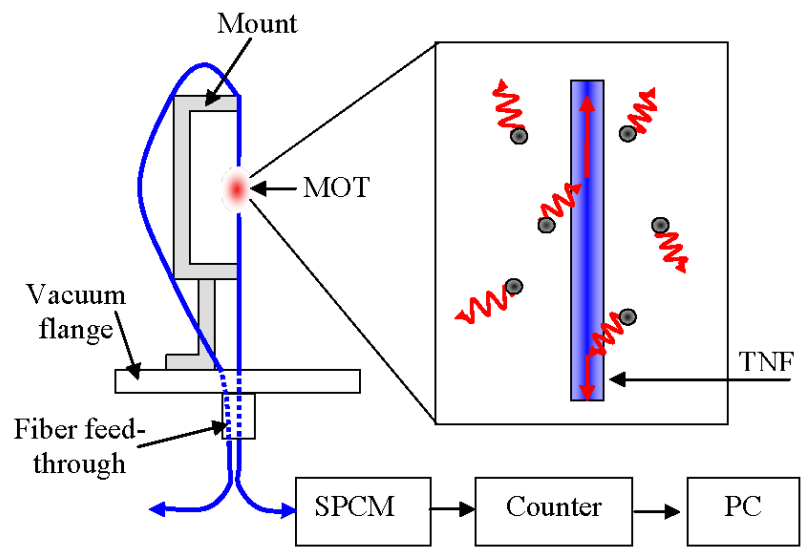

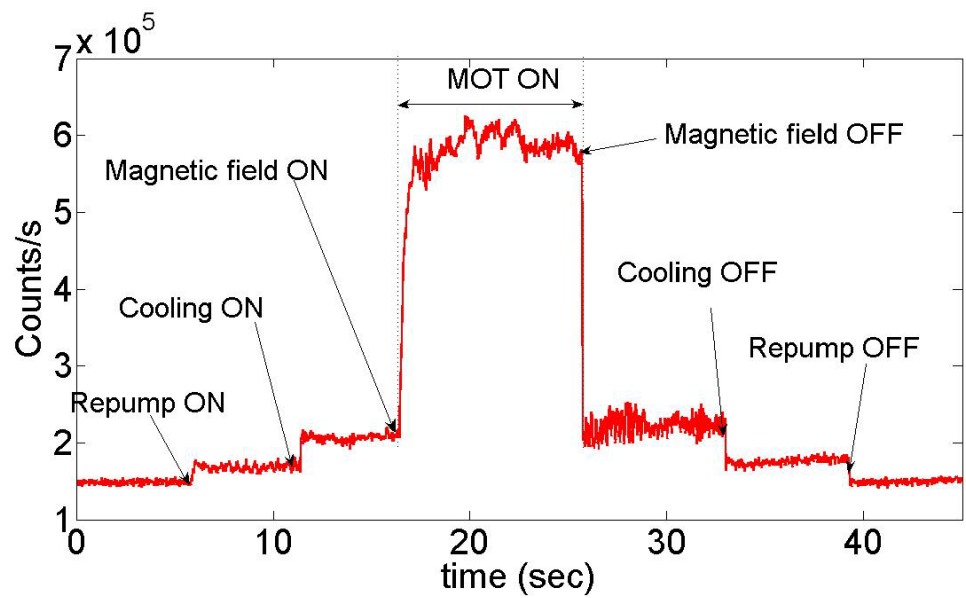

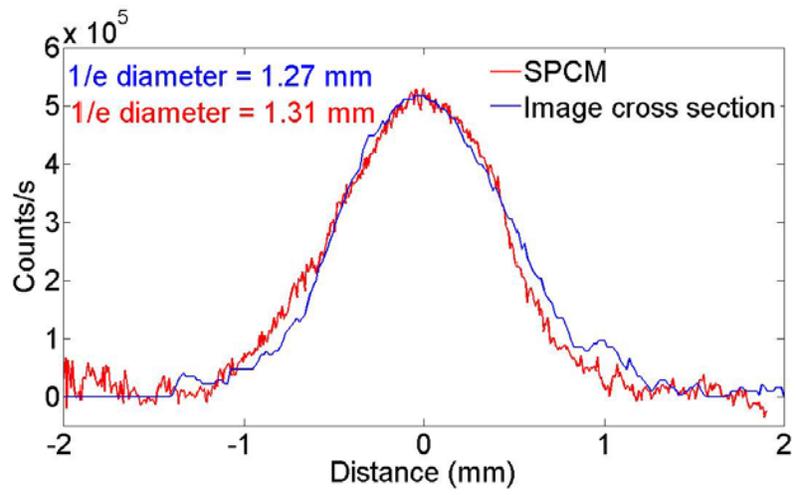

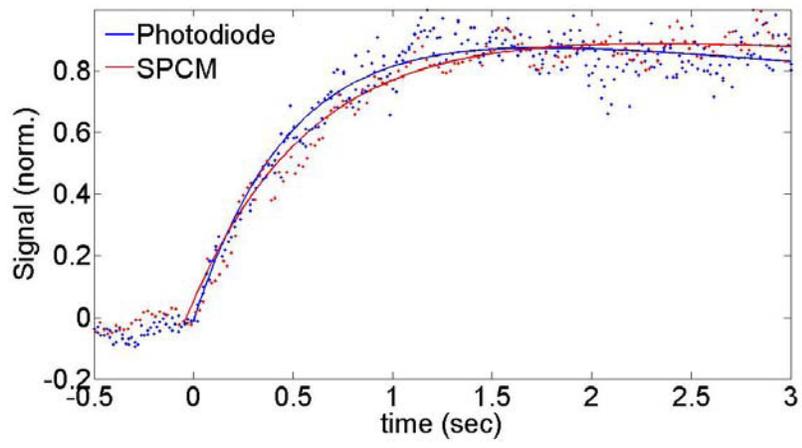

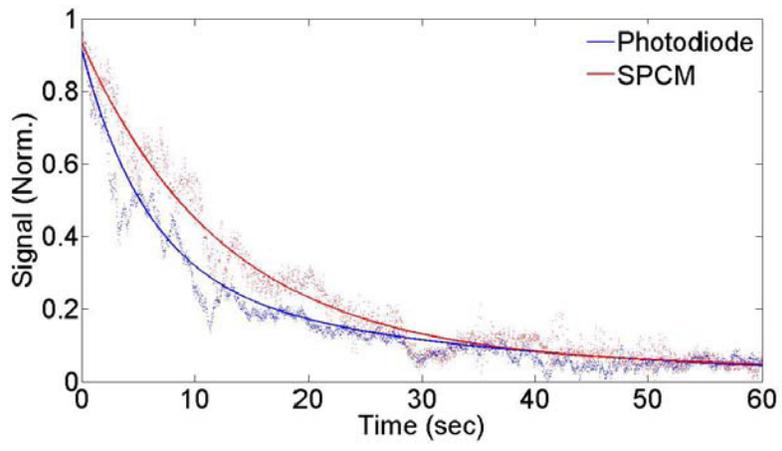